\documentclass[10pt,article]{article}
\usepackage[top=0.85in,left=0.75in,footskip=0.75in,marginparwidth=2in]{geometry}
\usepackage{physics}
\usepackage{nameref,hyperref}
\usepackage{graphicx}
\usepackage{color}
\usepackage{abstract,lipsum}
\usepackage{cite}

\usepackage[aboveskip=1pt,labelfont=bf,labelsep=period,singlelinecheck=off]{caption}

\makeatletter
\renewcommand{\@biblabel}[1]{\quad#1.}
\makeatother

\usepackage{lastpage,fancyhdr,graphicx}
\usepackage{epstopdf}

\usepackage{wrapfig}
\usepackage[pscoord]{eso-pic}
\usepackage[fulladjust]{marginnote}
\reversemarginpar

\begin{document}
\vspace*{0.35in}

\begin{flushleft}
{\LARGE
\textbf\newline{ Controllable optical bistability in a cavity optomechanical system with a Bose-Einstein condensate}
}
\newline
\\
Seyedeh Hamideh Kazemi\textsuperscript{1},
Saeed Ghanbari\textsuperscript{1},
Mohammad Mahmoudi\textsuperscript{1,*}
\\
\bigskip
\textsf{1} Department of Physics, University of Zanjan, University Blvd., 45371-38791, Zanjan, Iran
\\
* mahmoudi@znu.ac.ir

\end{flushleft}
\begin{abstract}
The optical bistability (OB) in a two-mode optomechanical system with a Bose-Einstein condensate (BEC) is studied. By investigating the behavior of steady state solutions, we show that how OB develops in the system for a certain range of cavity-pump detunings and pump amplitudes. We then investigate the effects of the decay rate of the cavity photons and coupling strength between the cavity and the BEC as well as the pump-atom detuning on the optical behaviour of the system. We find that one can control the OB threshold and width of the bistability curve via adjusting properly the decay rate, coupling strength and the detuning. By applying Routh-Hurwitz criterion, we then derive stability conditions for different branches of the OB curve. Moreover, by introducing an effective potential for the system, a simple physical interpretation is obtained.
\end{abstract}
\vspace*{0.2in}

\section{\label{sec:level1}Introduction}

In 1925, by generalizing the statistics of photons to a system of non-interacting and massive bosons \cite{2,3}, which had been developed by Bose \cite{1}, Einstein concluded that a finite fraction of the particles fill the same single-particle quantum state, below a critical temperature. This fascinating state of matter is now known as a Bose-Einstein condensate (BEC). After the experimental realization of Bose-Einstein condensation in dilute gases of rubidium \cite{4}, sodium \cite{5} and lithium \cite{6}, ultracold Bose gases became versatile and robust systems for probing different properties of quantum systems such as quantum turbulence \cite{Yukalov}, quantum information processing \cite{bloch} and entanglement \cite{tasgin}.

Superradiance originates from the famous argument by Dicke in 1954 \cite{dicke1} which is collective interaction of an ensemble of atoms with a single mode of radiation. Thereafter, collective interactions with fields have been a major subject of research over the past several years. In an experimental analysis in 2005 \cite{ottl}, a combined system with an ultracold atomic ensemble located in an optical cavity was reported and a few years later, strong coupling of a BEC to a cavity was realized \cite{brennecke,colombe}. Placing the BEC inside a cavity would offer several advantages. First, we can prepare the oscillator in its ground state without thermally activated excitations \cite{brennecke}. Second, it dramatically enhances the atom-light interaction by collective coupling to the same light mode \cite{colombe,black,gupta}. And third, decoherence effects associated with inhomogeneities and with atomic motions can be suppressed for a BEC \cite{sherson}. After this experimental realization, there has been a surge of interest in investigation of this system which leads to many interesting developments including cavity enhanced superradiance of a BEC \cite{salma}, self-organization of atoms \cite{nagy}, quantum information \cite{bohi} and superluminal light propagation \cite{kazemi}, to mention a few.

OB, predicted by Sz\"{o}ke and coworkers in 1969 \cite{szoke}, is a phenomenon in which optical resonators containing saturable absorbers have nonlinear characteristics and can exhibit hysteresis. It has been extensively demonstrated both experimentally and theoretically in many systems such as superconducting quantum circuits \cite{hamedi}, quantum dots \cite{mahmoudi2,hamedi2,mahmoudi} and quantum wells \cite{mahmoudi3,sete}. Recently, ultracold atom groups at Berkeley \cite{gupta} and  ETH Z\"{u}rich \cite{brennecke} have found OB in systems comprising of vapours of ultracold atoms trapped in optical cavities. Unlike an optomechanical system consisting of an empty cavity with a movable end mirror, the photon numbers in the cavity with a BEC are usually low and even below unity which is desirable for applications such as optical communication and quantum computation.

On the other hand, because of the practical applications of the OB such as all-optical switches, optical transistors and optical memory elements controlling the OB seems to be crucial for the prominent applications. However, from an experimental point of view due to lack of controllability, it has limited applications. In recent years, some attempts have appeared both theoretically and experimentally in order to control the OB. Joshi \textit{et al.} \cite{joshi}, manipulated the OB by changing the intensity and frequency detuning of the coupling field. Almost a year later, Chang and colleagues \cite{chang} showed controllable shift of the threshold points of the OB induced by two suitable tuned fields. By controlling this shift, all-optical flip-flop and storage of optical pulse signals were implemented. More recently, the OB of an ultracold atomic ensemble located in an optical cavity was investigated and it was shown that a transverse pumping field could be used to control the bistable behaviour \cite{yang}. In another study, the OB was investigated for BECs of atoms in a driven optical cavity with Kerr medium and it was shown that the OB can be controlled by adjusting Kerr interaction between the photons  \cite{kerr}. In this paper, we investigate effects of the system parameters, i.e., decay rate of the cavity photons, coupling strength between the cavity and the BEC as well as the pump-atom detuning, on the OB in an optomechanical system with a BEC. Controlling the OB is done by simply changing the corresponding parameters without necessity of another field or Kerr medium which can be much more efficient than the previous ones.

This paper is organized as follows. In section 2, we introduce the Hamiltonian of the system consisting of a BEC trapped in a cavity in the discrete-mode approximation. Then, in section 3, we focus on the conditions for bistability and controllability of the OB characteristics via adjusting properly the corresponding parameters of the system, which may lead to promising applications such as all-optical switching. We also investigate the stability of the different branches of the OB curve by applying Routh-Hurwitz criterion and find the minimum value of the threshold intensity. Moreover, we investigate the dynamic properties by introducing an effective potential for the system.

\section{Formalism of the Optomechanical System}
The system under consideration is depicted schematically in Fig. 1. We consider a cigar-shaped BEC of  $N$ $\rm{{}^{87}}$Rb atoms trapped in an optical ultrahigh-finesse Fabry-Perot cavity. The coupled BEC-cavity system is driven by a pump laser with frequency $\omega_{p}$ and amplitude $\vert  E_{p} \vert=\sqrt{\frac{2\kappa P_{p}}{\hbar\omega_{p}}}$ which is a function of laser power $P_{p}$ and decay rate of the cavity photons  $\kappa$. In the large detuning between the pump laser frequency and the atomic resonance, $\triangle_a=\omega_{p}-\omega_{a}$, we can eliminate the excited state of the atoms and the Hamiltonian describing the coupled BEC-cavity system reads \cite{maschler}
 \begin{eqnarray}\label{eq1}
\hat{H}_{eff}\ &=& \int \hat{\Psi}^{\dagger}({x}) \{\frac{-\hbar^{2}}{2 m} \frac{d^{2}}{dx^{2}}
+ V_{ext}({x})+\hbar U_{0} \cos^{2}(kx) \hat{a}^{\dagger} \hat{a} \}\hat{\Psi}({x})dx \nonumber\\
&+&\hat{H}_{A-A} + \hbar \Delta \hat{a}^{\dagger} \hat{a}+i \hbar E_{p} (\hat{a}^{\dagger}- \hat{a}).\
\end{eqnarray}
 \begin{figure}
 \hspace{+0.25cm}
\centerline{\includegraphics[width=8 cm]{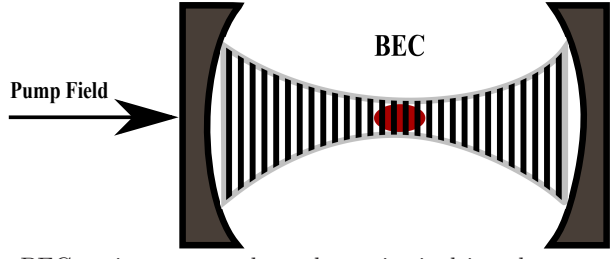}}
\caption{\small Sketch of a BEC-cavity system where the cavity is driven by a pump-laser field along the cavity axis.}
\label{figure 1}
\end{figure}
Here $\hat{a}^{\dagger} (\hat{a}) $ denotes the creation (annihilation) operator of the cavity photons with mode function  $\cos(kx)$ and wave vector $k=\frac{2\pi}{\lambda}$. The atomic field operator for the creation of an atom in the ground state with mass  $m$ at position $x$ is given by $\hat{\Psi}^{\dagger}({x})$. As seen in Eq. (\ref{eq1}), the atom-cavity photon interaction provides a dynamic and quantized potential lattice $U_0 \cos^{2}(kx) \hat{a}^{\dagger} \hat{a}$  for the atoms. Also, $U_{0}=\frac{g_{0}^2}{\Delta_a}$ denotes the potential depth for a single photon with the atom-photon coupling constant $g_0$. Cavity frequency is detuned from the pump laser field by $\triangle=\omega_{c}-\omega_{p}$. Moreover, $V_{ext}(x)$ and  $\hat{H}_{A-A}$ are the external trapping potential and  atom-atom interaction, respectively.
Applying the discrete-mode approximation \cite{zhang}, we can expand the atomic field operator as
\begin{equation}
\hat{\Psi}({x})= \varphi_{0} \hat{c}_{0}+ \varphi_{2}\hat{c}_{2}.
\end{equation}
Where $\varphi_{0}=1$ and  $\varphi_{2}=\sqrt{2}\cos(2kx)$ are two spatial modes  and $\hat{c}_{0}$ and $\hat{c}_{2}$ denote the  corresponding annihilation operators. For weakly interacting atoms in a shallow external trapping potential, after applying the Bogoliubov approximation and in a rotating frame, the Hamiltonian describing the coupled BEC-cavity reduces to
\begin{equation}\label{eq3}
\hat{H} = 4 \hbar\ \omega_{rec} \hat{c}_{2}^{\dagger} \hat{c}_{2} + \hbar \Delta_{c} \hat{a}^{\dagger} \hat{a} +\hbar g (\hat{c}_{2}^{\dagger} +\hat{c}_{2} ) \hat{a}^{\dagger} \hat{a} + i \hbar E_{p} (\hat{a}^{\dagger}- \hat{a}).
\end{equation}
Here, $\hbar \omega_{rec}=\frac{\hbar^2 k^2}{2 m}$ is the recoil energy. Also, $\Delta_c=\Delta+\frac{NU_{0}}{2} $ and  $g=\frac{U_0}{2} \sqrt{\frac{N}{2}}$  stand for the shifted cavity-pump detuning and the effective coupling strength between the cavity and the BEC, respectively. Moreover, the last term in the Hamiltonian, describes the classical pump light input.

From Eq. (\ref{eq3}), Langevin equations for the optomechanical system are given by
\begin{equation}
\frac{d}{dt} \hat{a}({t})=\\
 (- i \Delta_c - \frac{\kappa}{2})\ \hat{a}({t}) - i g \sqrt{2} \ \hat{X}({t}) \ \hat{a}({t})+E_{p},
\end{equation}
\begin{equation}\label{eq5}
\frac{d}{dt} \hat{X} ({t}) = 4\omega_{rec} \hat{P}({t}),
\end{equation}
\begin{equation}\label{eq6}
\frac{d}{dt} \hat{P}({t})=- 4 \omega_{rec}\ \hat{X}({t})-\gamma_{m}\hat{P}({t}) -  \sqrt{2} \ g \ \hat{a}^{\dagger}({t})\ \hat{a}({t}).
\end{equation}

 We have introduced the position and momentum operators of the BEC,  $\hat{X}=\frac{\hat{c}_{2}^{\dagger} +\hat{c}_{2}}{\sqrt{2}}$  and  $\hat{P}=\frac{\hat{c}_{2}-\hat{c}_{2}^{\dagger} }{ i\sqrt{2}}$  as well as the damping rate of the atomic excited state $\gamma_{m}$.

Assuming $\hat{a}({t})=\bar{a}+\delta\hat{a}({t})$, $ \hat{X}({t})=\bar{x}+\delta\hat{X}({t})$ and $ \hat{P}({t})=\delta\hat{P}({t})$, we first find the self-consistent steady state solutions
  \begin{equation}\label{eq7}
\bar{a} = \frac{E_{p}}{ i\Delta_c +i\ g \ \sqrt{2} \ \bar{x}+\frac{\kappa}{2}},
\end{equation}
\begin{equation}\label{eq8}
\bar{x} = \frac{- g \sqrt{2} |\bar{a}|^{2}}{4\ \omega_{rec}},
\phantom{\hspace{1.5cm}}
\end{equation}
 here
 $\bar{a}$  and $\bar{x}$ stand for the interactivity field and the mechanical position. Moreover, $\delta \hat{a}$ and $\delta \hat{X}$  are the fluctuations around these steady states. Combining Eqs. (\ref{eq7}) and (\ref{eq8}), leads
\begin{equation}\label{eq9}
n=\frac{E_{p}^2}{(\frac{\kappa}{2})^2+(\Delta_c-\frac{g^2}{2 \omega_{m}} n)^2},
\end{equation}
where $\omega_{m}=4 \omega_{rec}$ and $n=|\bar{a}|^{2}$ is the number of photons.
Rearranging Eq. (\ref{eq9}), the resulting equation is
 \begin{equation}\label{eq10}
n((\frac{\kappa}{2})^2+(\Delta_c-\frac{g^2}{2 \omega_{m}} n )^2)=E_{p}^2,
\end{equation}
Setting Eq. (\ref{eq10}) to zero, gives a third-order polynomial equation of the form
\begin{equation}\label{eq14}
n^3+a_2n^2+a_1n+a_0=0,
\end{equation}
where
\begin{equation}
b=\frac{g^4}{4 \omega ^2_{m}}, \ \ a_2=\frac{-4 \Delta_{c} \omega_m}{g^2},\ \ a_1=\frac{\frac{ \kappa^2}{4}+ \Delta ^2 _{c}}{b}, \ \ a_0=\frac{-E_{p}^2}{b}.
\end{equation}
By noting the fact that the steady state photon numbers exhibit OB for a certain range of cavity-pump detuning by increasing the amplitude of the pump field, in the following we are going to calculate the critical values of cavity-pump detuning and pump amplitude. So the system first becomes bistable at a single value of the detuning, denoted by $\Delta_{cr}$ and the critical pump amplitude, at which the bistability at $\Delta_{cr}$  occurs, is denoted by $E_{cr}$.
In order to calculate the critical values, let us have a look  at Eq. (\ref{eq14}). The critical points of the third-order polynomial equation are obtained by setting derivative of the equation equal to zero and are given by $n=\frac{-a_2\pm\sqrt{a_2^2-3a_1}}{3}$. In the case where $\sqrt{a_2^2-3a_1}$  is positive, the function has a local minimum and a local maximum. Equating  $\sqrt{a_2^2-3a_1}$ to zero so that the function has only a critical point, the critical cavity-pump detuning is given by
\begin{equation}
\Delta_{cr}=\frac{\sqrt{3}}{2}\kappa,
\end{equation}
and we define the critical pump amplitude, at which the bistability at $\Delta_{cr}$ occurs, as
\begin{equation}
E_{cr}=\sqrt{\frac{\omega_m  \kappa^3}{6\sqrt{3}\ g^2}}.
\end{equation}
In the following, by using the ansatz and retaining only first order terms in the small quantities $\delta \hat{a},\delta \hat{X}$, $ \delta \hat{a}^{\dagger}$ and $\delta \hat{P}$, one gets the linearized quantum Langevin equations for the fluctuation operators
 \begin{equation}\label{6}
\frac{d}{dt}\delta \hat{a}({t})=\\
 (- i \bar{\Delta}_c - \frac{\kappa}{2})\delta \hat{a}({t}) - i g \sqrt{2} \ \bar{a} \ \delta\hat{X}({t}) ,
 \phantom{\hspace{0.75cm}}
\end{equation}
\begin{equation}\label{7}
\frac{d}{dt} \delta\hat{X}({t})= \omega_{m} \delta\hat{P}({t}),
\end{equation}
 \begin{equation}\label{8}
\frac{d}{dt} \delta\hat{P}({t})=- \omega_{m}\delta\hat{X}({t}) - \gamma_{m}\delta\hat{P}({t})-  \sqrt{2} g (\bar{a}\  \delta\hat{a}^{\dagger}({t}) + \bar{a} ^{*}\delta\hat{a}({t})),
\end{equation}
where $\bar{\Delta}_c=\Delta_c+\ g \ \sqrt{2} \ \bar{x}  $.
\section{Results and discussion}

\begin{figure*}
\centerline{\includegraphics[width=12cm]{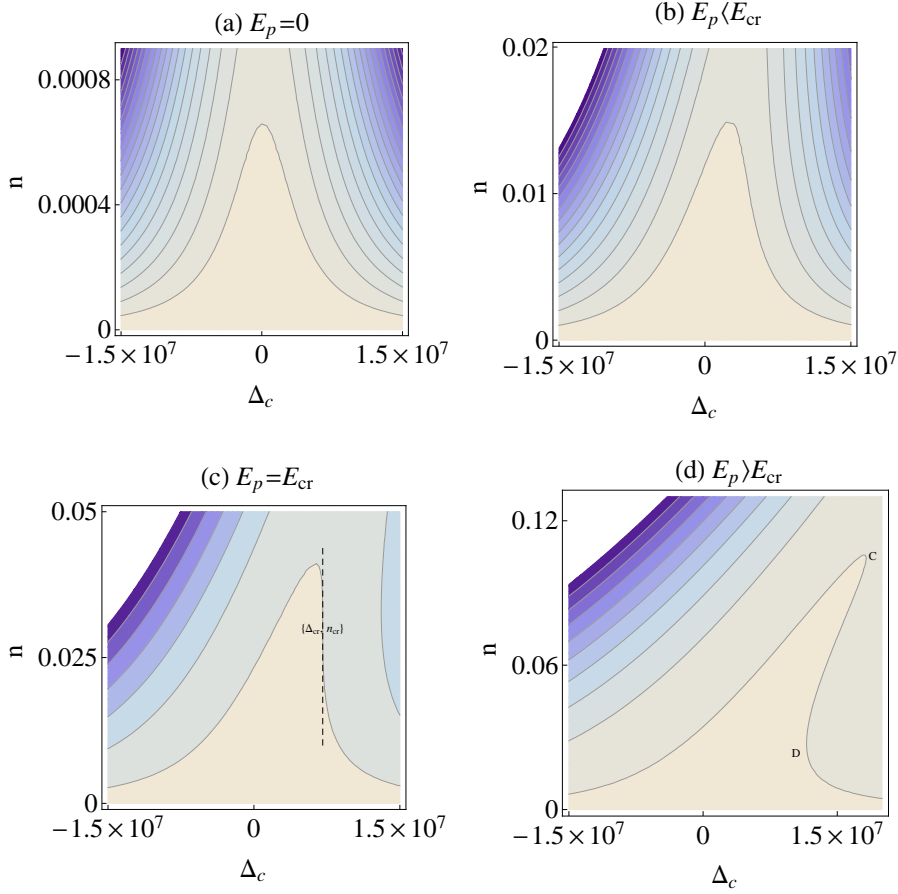}}
\caption{\small steady state photon numbers as a function of the shifted cavity-pump detuning for the four different amplitude of the pump field, $E_{p}=0$ (a), $E_{p}=0.5 $ MHz (b), $E_{p}=0.83 $ MHz (c) and $E_{p}=1.5 $ MHz (d). Other used parameters are $\kappa=2\pi\times 1.3$ KHz, $ N=1.2\times10^5$, $\Delta_a=2\pi\times 32 $ GHz, $g_0=2\pi\times 10.9 $ MHz and $\omega_{m}=2\pi\times 15.2 $ KHz .}
\end{figure*}

In the begining of this section, we present the results based on the numerical solutions of Eq. (\ref{eq9}) and focus on the conditions for bistability. 
\begin{figure}[!ht]
\centerline{\includegraphics[width=12cm]{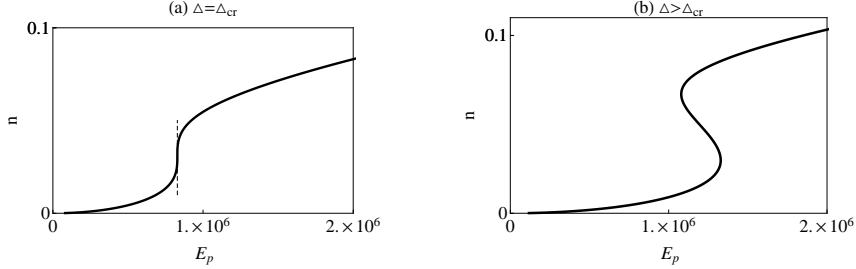}}
\caption{\small S-shaped curve for the steady state photon numbers as a function of the amplitude of the pump for $\Delta_{c}= 7 $ MHz (a) and $\Delta_{c}= 11 $ MHz (b). Other parameters are the same as in Fig. 2.}
\end{figure}
In Fig. 2, we show how the photon numbers depend on the shifted cavity-pump detuning for given amplitudes of the pump $E_{p}$. Experimentally used parameters are $\kappa=2\pi\times 1.3 $ MHz, $ N=1.2\times10^5$, $\Delta_a=2\pi\times 32$ GHz, $g_0=2\pi\times 10.9$ MHz  and $\omega_{m}=2\pi\times 15.2$ KHz \cite{brennecke}. It can be seen from Eq. (\ref{eq9}) that when $E_{p}$ is sufficiently small, the photon numbers are also small, so we can neglect  $ n^2$ and we have a symmetrical lorentzian curve centred at $\Delta_c=0$. In Fig. 2(b), it is shown that by increasing the amplitude of the pump field, the lorentzian curve is largely unchanged; however, the curve becomes more and more asymmetric and its only maximum moves to the right. But as $E_{p}$ reaches a critical value, the nature of the curve changes so that it has an infinite slope at  $\Delta=\Delta_{cr}$, according to Fig. 2(c). For amplitudes beyond the critical value, Eq. (\ref{eq9}) has three real roots, corresponding to the branch CD in Fig. 2(d) where the multiple solutions indicate the bistable behaviour.

The bistable behaviour can also be seen from the OB hysteresis curve, shown in Fig. 3. For the detuning above the critical value by increasing the amplitude of the pump field system goes to the bistability branch, according to Fig. 3(b). Consider that amplitude increases from zero gradually; the photon numbers initially follow the lower stable branch. When the amplitude increases to the first bistable point, it jumps to the upper stable branch and if the amplitude is increased further, continues to follow that branch. Now if we start decreasing the amplitude, the steady state photon numbers remain on the upper stable branch at first; however, by decreasing the amplitude even further, when it reaches the second bistable point, it jumps down to the lower stable branch and continues to decrease along that branch for further decrease of the amplitude. We Note that placing the BEC inside an optical cavity allows for OB at extremely low cavity photon numbers; however, an empty optomechanical system does not exhibit the behaviour.

In the following, we are mainly focused on controllability of the OB. 
\begin{figure}[!hb]
\centerline{\includegraphics[width=8cm]{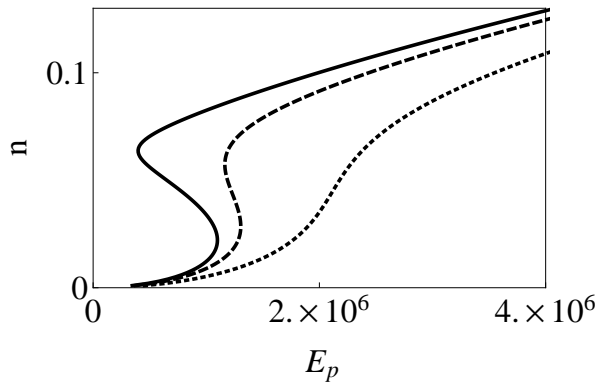}}
\caption{\small Steady state photon numbers as a function of the amplitude of the pump for different decay rate of the cavity photons $ \kappa=2 \pi \times 0.5$ MHz (solid line), $\kappa=2\pi \times 1.5$ MHz (dashed line) and $\kappa=2\pi \times 3$ MHz (dotted line). Other parameters are the same as in Fig. 3(b).}
\end{figure}
First, we are going to examine the effects of the decay rate of the cavity photons on the OB. Figure 4 depicts steady state photon numbers as a function of amplitude of the pump for various decay rates. Here, we can observe that by increasing the decay rate, width of the bistability curve decreases. In addition, the OB threshold increase by increasing the value of the decay rate. By further increasing the decay rate, the OB may be eliminated. In order to clarify the behaviour of the OB curve with variation of the decay rate, we present the Fig. 5 considering the same parameters used for Fig. 3(b). As one expects, decay rate of the cavity photons pushes the OB threshold to higher values.

\begin{figure}
\centerline{\includegraphics[width=8cm]{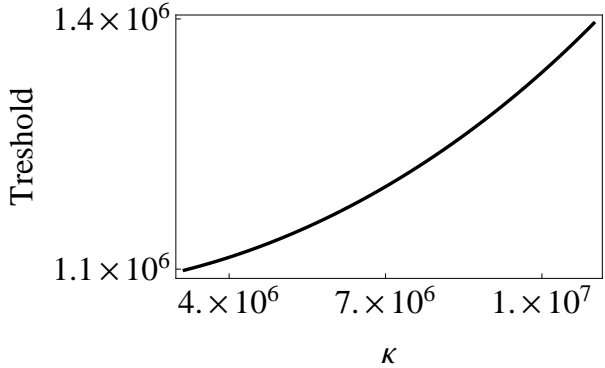}}
\caption{\small Threshold intensity as a function of $ \kappa $. Other parameters are the same as in Fig. 3(b).}
\end{figure}

In what follows, we will investigate the effects of the coupling strength between the cavity and the BEC on the OB threshold and the width of the bistability curve. In refrence \cite{colombe}, the authors have shown the possibility of controlling the coupling strength by locating the BEC anywhere within the cavity, so we can introduce another control parameter. Steady state photon numbers as a function of the amplitude of the pump and for different coupling strengths are plotted in Fig. 6. Other parameters are the same as in Fig. 3(b). As is seen, increasing the coupling strength, i.e.,  $g_{0}$, causes the system to become bistable at lower amplitudes of the pump.  It also decreases the width of the bistability curve. Therefore, it is considered as another control parameter which can provide the possibility of realizing a controllable optical switch. Moreover, the coupling strength allows for the OB at extremely low photon numbers as well as fast and low-threshold optical switches.

Then we show that optical behaviour is sensitive to the detuning between the pump laser field and the atomic resonance.
\begin{figure}[!hb]
\centerline{\includegraphics[width=8cm]{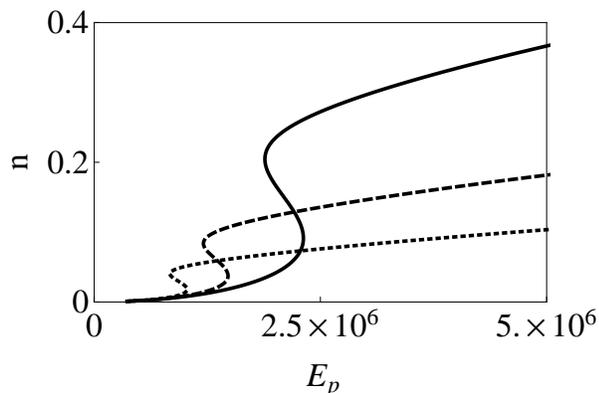}}
\caption{\small Steady state photon numbers as a function of the amplitude of the pump for three different coupling strength $ g_{0}=2\pi\times 8 $ MHz (solid line), $g_{0}=2\pi\times 10 $ MHz (dashed line) and $g_{0}=2\pi\times 12 $ MHz (dotted line). The value of other parameters is the same as those in Fig. 3(b).}
\end{figure}
Figure 7 depicts steady state photon numbers versus the amplitude of the pump for $\Delta_a =2 \pi\times 30$ GHz (solid line), $\Delta_a =2 \pi\times 40 $ GHz (dotted line) and $ \Delta_a =2 \pi\times 50 $ GHz (dashed line). According to the figure, decreasing the pump-atom detuning causes the system to become bistable at lower amplitudes and also decrease the width of the bistability curve. This behavior opens the possibility of the optical switching by varying the pump laser field frequency.

As the OB threshold intensity is a function of the corresponding parameters, we minimize it over the range of experimental data \cite{brennecke,colombe} and find the minimum threshold intensity of the OB, $E_{p_{min}}=1 $ MHz and the corresponding laser power $P_{p_{min}}=3\times 10^{-14} $W.

\begin{figure}
\centering
\includegraphics[width=8cm]{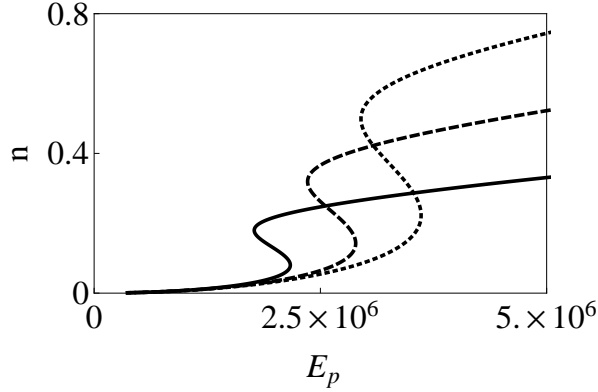}
\caption{\small Steady state photon numbers as a function of the amplitude of the pump for $\Delta_a =2 \pi\times 30 $ GHz (solid line), $\Delta_a =2 \pi\times 40$ GHz (dashed line) and $ \Delta_a =2 \pi\times 50 $ GHz (dotted line). Other parameters are the same as in Fig. 3(b).}
\end{figure}

In the following, we are going to study the stability of the different branches of the OB curve. Applying the Routh-Hurwitz criterion  to the Eqs. (\ref{6}), (\ref{7}) and (\ref{8}), we can get the stability conditions

\begin{equation}
\begin{split}
s_1 &= \kappa \gamma_{m} \lbrace [\frac{\kappa^2}{4} +(\omega_{m}-\bar{\Delta}_c)^2][ \frac{\kappa^{2}}{4}+(\omega_{m}+\bar{\Delta}_c)^2]\\
 & + \gamma_{m} [ (\gamma_{m}+\kappa) (\frac{\kappa^{2}}{4}+\bar{\Delta}_c^{2})+\kappa \omega^{2}_{m}] \rbrace +4 \,g^2 \,n\,\omega_{m} \,\bar{\Delta}_c (\gamma_{m}+\kappa)^2>0,
\end{split}
\end{equation}
\begin{equation}\label{eq17}
s_2=\omega_{m} (\frac{\kappa^{2}}{4}+\bar{\Delta}_{c}^{2})-4\,g^2\, n\, \bar{\Delta}_c >0.\phantom{\hspace{1.25cm}}
\end{equation}
For the positive detuning, $\bar{\Delta}_c>0$, the first condition is always satisfied. In Fig. 8, we see that for the middle branch, $s_2$ is negative and so the steady states on the region are unstable. It is worth mentioning that the condition is consistent with the fact that the branches with  negative slope are always unstable \cite{ikeda}.
\begin{figure*}[!hb]
\centerline{\includegraphics[width=12cm]{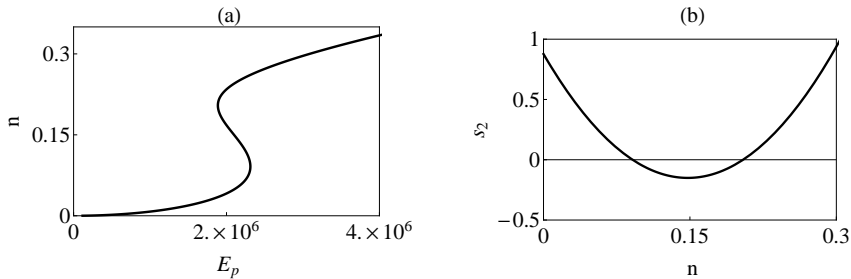}}
\caption{\small (a) Behavior of steady state photon numbers in response to the pump field. (b) shows the normalized second stability condition $s_2$ versus the photon numbers. Parameters values are the same as in Fig. 3(b), with except that $g_{0}$ is replaced by the coupling strength of Fig. 6(a). }
\end{figure*}

In what follows, we will try to obtain the physical interpretation of the condition. Note that experimentally \cite{brennecke,colombe}, the decay rate of the cavity photons is often two orders of magnitude faster than the mechanical motion of the condensate, so we safely assume that the cavity field follows the condensate adiabatically. Setting $\frac{d}{dt}\delta\hat{a} =0$ in Eq. (\ref{6}), one gets:
\begin{equation}
\delta{a({t})=- \frac{ig\sqrt{2}\  \bar{a} \ \delta{X({t})}} {i\bar{\Delta}_{c}+\frac{\kappa}{2}}},
\phantom{\hspace{4.75cm}}
\end{equation}
\begin{equation}
\delta{a^{*}({t})=\frac{ig\sqrt{2} \ \bar{a}^{*} \ \delta{X({t})}} {-i\bar{\Delta}_{c}+\frac{\kappa}{2}}},
\phantom{\hspace{4.5cm}}
\end{equation}
\begin{equation}\label{9}
\frac{d}{dt} \delta\hat{P}({t})=-(\omega_{m}+\frac{4 g^{2}n \bar{\Delta}_{c}}{\bar{\Delta}_{c}^{2}+\frac{\kappa^{2}}{4}})  \delta\hat{X}({t})- \gamma_{m}\delta\hat{P}({t}),
\phantom{\hspace{1cm}}
\end{equation}
\begin{equation}\label{19}
\frac{d^{2}}{dt^{2}} \delta\hat{X}({t})  + \gamma_{m}\frac{d}{dt}\delta\hat{X}({t})  + (\omega^{2}_{m}-\frac{4 g^{2}\ n \omega_{m} \bar{\Delta}_{c}}{\bar{\Delta}_{c}^{2}+\frac{\kappa^{2}}{4}})  \delta\hat{X}({t})=0.
\end{equation}

It can be seen that the coefficient of $\delta\hat{X} $ in third term, corresponds to a restoring force. On the other hand, from the second condition, Eq. (\ref{eq17}), we can see that the mechanical oscillator is stable if the restoring force is positive.  In the middle branch, the third coefficient in the Eq. (\ref{19}) is negative and so it is no longer a restoring force leading to the unstable behaviour. The phenomenon can be also viewed as optical spring effect \cite{genes,aldana}. It is worth to mention that except here, we do not make an adiabatic approximation.

Considering that OB can be well understood in the framework of the simple double-well potential model, so we proceed to obtaining the effective potential. Putting the interactivity field into Eqs. (\ref{eq5}) and (\ref{eq6}), we get the evolution of the displacement for the BEC
\begin{equation}
\frac{d^{2}}{dt^{2}}X({t})  = -\omega^{2}_{m}X({t})-\sqrt{2}\ g \ \omega_{m} n,
\end{equation}
and the potential is given by
\begin{equation}
V(X)= \dfrac{1}{2} \omega^{2}_{m} X^{2} +\frac{E^2_{p}\, \arctan(\frac{\Delta_c + \sqrt{2} \, g\, X}{\sqrt{1/4} \kappa})}{\sqrt{4} \kappa}.
\end{equation}
As is seen, we have a sum of the harmonic potential and inverse tangent terms. The latter may lead to two minima and the double-well like potential.
\begin{figure}
\centerline{\includegraphics[width=8cm]{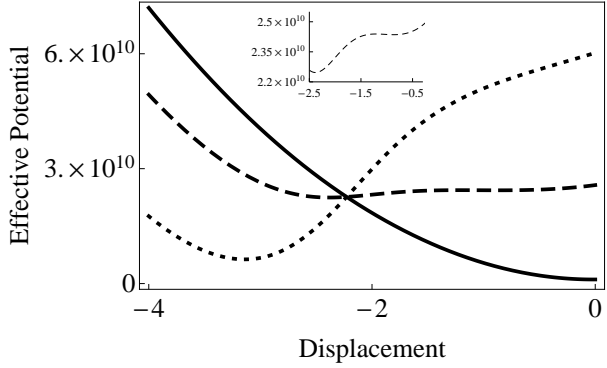}}
\caption{\small The effective potential for the different pump amplitude. The values are $E_{p}=200 $ KHz (solid line), $E_{p}=980 $ KHz (dashed line) and  $E_{p}=2 $ MHz (dotted line). The top and the bottom curve have only one minimum, while the middle one has two minimum and one maximum. The inset shows a zoom-in of the middle curve. Parameters values are the same as in Fig. 3(b).}
\end{figure}

Next, we investigate the dynamic properties by the effective potential which is plotted in Fig. 9 as a function of the displacement and for different pump amplitudes. The potential with $E_{p}=200 $ KHz is shown as the solid line. It gets the form of  the harmonic potential and has a single minimum close to the origin, corresponding to the stable lower branch. The dashed line corresponds to the case $E_{p}=980 $ KHz which lies in the OB region.
So, the potential is a double-well with three extrema, two minimum and one maximum. The first minimum located in the vicinity of the origin, corresponding to the lower branch of OB and the other one, is far away from the origin and corresponds to the upper branch. The maximum also represents the unstable branch. For  $E_{p}=2 $ MHz, the inverse tangent term gives just a shift to the potential and so it has a new minimum at a larger displacement corresponding to the purely upper stable branch.

\section{Conclusion}
We investigated the OB in a system consisting of a BEC inside an optical cavity. It was first shown that how the bistability develops in this cavity optomechanical system for a certain range of cavity-pump detunings and pump amplitudes. Further, it was demonstrated that one can control the OB threshold and the width of the bistability curve by means of the decay rate of the cavity photons, coupling strength between the cavity and the BEC as well as the pump-atom detuning. To the best of our knowledge, our work is the first of its kind to investigate the effects of the system parameters on the optical behaviour. Another additional advantage is that extremely low cavity photon numbers exhibit bistable behaviour which is absent for previous systems. Also, the system parameters allow for a considerable reduction of the threshold and this phenomenon has potential applications in optical switches at very low intensities. Moreover, width of the bistability curve may be altered by changing the parameters allowing all-optical flip-flop and storage of optical pulse signals. We then analyzed the stability of the different branches of the OB curve by applying the Routh-Hurwitz criterion as well as offering a physical interpretation for these conditions. Finally, we derived an effective potential in order to have an intuitive picture.
\newline

\bibliographystyle{}

\end{document}